  \documentclass[preprint]{aastex61}
\usepackage{longtable}
\accepted{in ApJL}

\shorttitle{Characterization of the Ross 128 Exoplanetary System}
\shortauthors{Souto et al.}

\begin{document}

\title{Stellar and Planetary Characterization of the Ross 128 Exoplanetary System from APOGEE Spectra}

\correspondingauthor{Diogo Souto}
\email{souto@on.br, diogodusouto@gmail.com}

\author[0000-0002-7883-5425]{Diogo Souto}
\affiliation{Observat\'orio Nacional, Rua General Jos\'e Cristino, 77, 20921-400 S\~ao Crist\'ov\~ao, Rio de Janeiro, RJ, Brazil}

\author[0000-0001-8991-3110]{Cayman T. Unterborn}
\affiliation{School of Earth and Space Exploration, Arizona State University, Tempe, AZ 85287, USA}

\author{Verne V. Smith}
\affiliation{National Optical Astronomy Observatory, 950 North Cherry Avenue, Tucson, AZ 85719, USA}

\author{Katia Cunha}
\affiliation{Observat\'orio Nacional, Rua General Jos\'e Cristino, 77, 20921-400 S\~ao Crist\'ov\~ao, Rio de Janeiro, RJ, Brazil}
\affiliation{Steward Observatory, University of Arizona, 933 North Cherry Avenue, Tucson, AZ 85721-0065, USA}

\author{Johanna Teske}
\affiliation{Department of Terrestrial Magnetism, Carnegie Institution for Science, Washington, DC 20015}
\affiliation{The Observatories of the Carnegie Institution for Science, 813 Santa Barbara Street, Pasadena CA, 91101.}
\affiliation{Hubble Fellow}

\author{Kevin Covey}
\affiliation{Department of Physics \& Astronomy, Western Washington University, Bellingham, WA, 98225, USA}

\author{B\'arbara Rojas-Ayala}
\affiliation{Departamento de Ciencias Fisicas, Universidad Andres Bello, Fernandez Concha 700, Las Condes, Santiago, Chile}

\author{D. A. Garc\'ia-Hern\'andez}
\affiliation{Instituto de Astrof\'isica de Canarias, E-38205 La Laguna, Tenerife, Spain}
\affiliation{Departamento de Astrof\'isica, Universidad de La Laguna, E-38206 La Laguna, Tenerife, Spain}

\author{Keivan Stassun}
\author[0000-0002-3481-9052]{Keivan G. Stassun}
\affiliation{Department of Physics and Astronomy, Vanderbilt University, 6301 Stevenson Center Ln., Nashville, TN 37235, USA
Department of Physics, Fisk University, 1000 17th Ave. N., Nashville, TN 37208, USA}


\author{Olga Zamora}
\affiliation{Instituto de Astrof\'isica de Canarias, E-38205 La Laguna, Tenerife, Spain}
\affiliation{Departamento de Astrof\'isica, Universidad de La Laguna, E-38206 La Laguna, Tenerife, Spain}

\author{Thomas Masseron}
\affiliation{Instituto de Astrof\'isica de Canarias, E-38205 La Laguna, Tenerife, Spain}
\affiliation{Departamento de Astrof\'isica, Universidad de La Laguna, E-38206 La Laguna, Tenerife, Spain}

\author{J. A. Johnson}
\affiliation{Department of Astronomy, The Ohio State University, Columbus, OH 43210, USA}

\author{Steven R. Majewski}
\affiliation{Department of Astronomy, University of Virginia, Charlottesville, VA 22904-4325, USA}

\author[0000-0002-4912-8609]{Henrik J\"onsson}
\affiliation{Lund Observatory, Department of Astronomy and Theoretical Physics, Lund University, Box 43, SE-221 00 Lund, Sweden}

\author{Steven Gilhool}
\affiliation{Department of Physics and Astronomy, University of Pennsylvania, 209 S. 33rd Street, Philadelphia, PA 19104}

\author{Cullen Blake}
\affiliation{Department of Physics and Astronomy, University of Pennsylvania, 209 S. 33rd Street, Philadelphia, PA 19104}

\author{Felipe Santana}
\affiliation{Universidad de Chile, Av. Libertador Bernardo O’Higgins 1058, Santiago de Chile}

\begin{abstract}

The first detailed chemical abundance analysis of the M dwarf (M4.0) exoplanet-hosting star Ross 128 is presented here, based upon near-infrared (1.5--1.7 \micron) high-resolution ($R$$\sim$22,500) spectra from the SDSS-APOGEE survey.
We determined precise atmospheric parameters $T_{\rm eff}$=3231$\pm$100K, log$g$=4.96$\pm$0.11 dex and chemical abundances of eight elements (C, O, Mg, Al, K, Ca, Ti, and Fe), finding Ross 128 to have near solar metallicity ([Fe/H] = +0.03$\pm$0.09 dex).
The derived results were obtained via spectral synthesis (1-D LTE) adopting both MARCS and PHOENIX model atmospheres; stellar parameters and chemical abundances derived from the different adopted models do not show significant offsets.
Mass-radius modeling of Ross 128b indicate that it lies below the pure rock composition curve, suggesting that it contains a mixture of rock and iron, with the relative amounts of each set by the ratio of Fe/Mg. 
If Ross 128b formed with a sub-solar Si abundance, and assuming the planet's composition matches that of the host-star, it likely has a larger core size relative to the Earth despite this producing a planet with a Si/Mg abundance ratio $\sim$34\% greater than the Sun.
The derived planetary parameters -- insolation flux (S$_{\rm Earth}$=1.79$\pm$0.26) and equilibrium temperature ($T_{\rm eq}$=294$\pm$10K) -- support previous findings that Ross 128b is a temperate exoplanet in the inner edge of the habitable zone.

\end{abstract}


\section{Introduction}

Nearby M dwarfs likely provide some of the best opportunities for detecting and characterizing potentially ``Earth-like" exoplanets in the near future.
M dwarfs produce larger observational signatures from low-mass exoplanets through both the radial velocity and transit methods (\citealt{Shields2016}, \citealt{Charbonneau2007}), making it easier to discover Earth-size or Earth-mass exoplanets orbiting these stars.
Discoveries of Earth-size or Earth-mass exoplanets have become common around low-mass stars thanks to the efforts of high cadence radial velocity ($RV$) programs, as well as the Kepler mission (\citealt{Batalha2013}). 
Proxima Cen b and the recent announcement of Ross 128b are examples of RV-detected exoplanets (\citealt{Anglada-Escud2016}, \citealt{Bonfils2017}), while TRAPPIST-1, Kepler-138, and Kepler-186 (\citealt{Gillon2017}, \citealt{JontofHutter2015}, \citealt{Quintana2014}, \citealt{Souto2017}) are examples of transiting exoplanet systems with cool M dwarf host stars.


An approach to study the exoplanet composition, albeit an indirect one, is the analysis of the individual host star.
This method can, for example, provide measurements of a star's C and O abundance, which play a role in the ice and gas chemistry in protoplanetary disks, as well as Mg, Fe, and Si abundances, that potentially control a rocky planet's core to mantle mass ratios (\citealt{Bond2010}, \citealt{DelgadoMena2010}, \citealt{Thiabaud2015}, \citealt{Dorn2017}, \citealt{Santos2017}, and \citealt{UnterbornPanero2017}).
Until recently, detailed abundance measurements of M dwarfs were lacking, due in part to the difficulty of obtaining high-S/N, high-resolution spectra, as well as strong molecular absorption from species such as TiO in the optical or H$_{2}$O in the near-infrared (\citealt{Allard2013}). 

The previous work of \cite{Souto2017} demonstrated that effective temperatures and detailed individual abundances of 13 elements can be measured from near-infrared (NIR) H-band high-resolution APOGEE (Apache Point Galactic Evolution Experiment; \citealt{Majewski2017}) spectra of warm M dwarfs ($T_{\rm eff}$ $\sim$3900K; see also \citealt{Schmidt2016}). Also using high-resolution spectra, \cite{Onehag2012}, \cite{Lindgren2017}, have shown that stellar metallicities can be studied from J-band spectra.
Most studies, however, use photometric calibrations to determine M dwarf stellar parameters e.g., $T_{\rm eff}$, log$g$, and mass (\citealt{Delfosse2000}, \citealt{Bonfils2005}, \citealt{Mann2015}) and low-resolution NIR spectroscopy to determine M-dwarf metallicities ([Fe/H]) from equivalent widths in the K-band (\citealt{RojasAyala2012}). 
These methods provide a good estimate of the stellar parameters and metallicity; in the context of exoplanet studies, accurate values for $T_{\rm eff}$ and $R_{\star}$ are needed in order to better constrain the exoplanet properties (e.g. insolation, equilibrium temperature). 
In addition, precise host star abundance measurements for particular elements can help to constrain the composition of the initial refractory materials that build rocky exoplanets.

In this work, we perform the first detailed abundance analysis for Ross 128, a cool M dwarf (M4.0) exoplanet host.  We use $R \sim$22,500 H-band APOGEE spectra to derive precise atmospheric parameters and chemical abundances of eight elements adopting a similar methodology as \cite{Souto2017}, but extending the analysis to the cool M dwarf regime.
Such parameters are needed to better characterize the mass ($Msin(i)$ $\sim$1.35 M$_{\oplus}$) of the 9.9 day period exoplanet around Ross 128, the second closest (3.37 pc) terrestrial-mass planet, recently discovered by \cite{Bonfils2017}.

\section{Observations and Spectrum Synthesis}

The APOGEE survey (\citealt{Majewski2017}, see also \citealt{Blanton2017}, \citealt{Zasowski2017}) observes primarily red giants, but has also observed M-dwarfs in the solar neighborhood to fill missing fibers or as part of ancillary projects. 
With APOGEE NIR spectral coverage ($\lambda$15,150-17,000\AA) and high-resolution spectroscopic capabilities (\citealt{Gunn2006}, \citealt{Wilson2010}), the APOGEE spectrograph has turned out to be an excellent instrument for detailed studies of M dwarfs (see \citealt{Souto2017}).

The APOGEE spectra of Ross 128 (HIP 57548; GJ 447; 2M11474440+0048164) were obtained with the fiber feed to the NMSU 1.0 meter telescope at APO.
We use the processed spectrum of Ross 128 from the DR14 pipeline (\citealt{DR14}, \citealt{Nidever2015}) representing the combination of two individual observations both obtained on 28 January 2014. 
The resulting S/N (per-pixel) in the combined spectrum is 230.

As a first step in the analysis we conducted a careful identification of the main spectral features in the APOGEE spectrum of Ross 128, given that this is a much cooler M dwarf than those previously analyzed in \cite{Souto2017}. 
Figure 1 displays the molecular and atomic lines identified; in this $T_{\rm eff}$ regime most of the APOGEE spectrum is dominated by H$_{2}$O or FeH lines but some atomic lines of Fe I, Mg I, Ca I, and Al I are also seen. We performed a line-by-line manual abundance analysis to determine the atmospheric parameters of Ross 128 (Section 2.1) and individual chemical abundances of the elements C, O, Mg, Al, K, Ca, Ti, and Fe.
The adopted transitions in the line-by-line abundance analysis (a total of 86 lines) are indicated with black vertical tick marks in Figure 1. 
Figure 1 also shows the best fitted synthetic spectra obtained from this analysis, as discussed below.

A spectrum synthesis analysis is required to analyze cool star (M-dwarf) spectra, rather than the equivalent-width method, due to molecular blends that blanket their spectra.
In this work, we adopted 1-D plane-parallel LTE models from MARCS and BT-Settl PHOENIX (\citealt{Gustafsson2008}, \citealt{Allard2013}). The synthetic spectra were computed with the Turbospectrum code (\citealt{AlvarezPLez1998}, \citealt{Plez2012}) using the modified version of the APOGEE line list (\citealt{Souto2017}, \citealt{Shetrone2015}) that takes into account transitions of FeH. 
The synthetic spectra were broadened with a Gaussian profile corresponding to the APOGEE resolution (a full width half maximum, FWHM, $\sim$0.73\AA{}).
We derived a vsin$i$$\leq$8 km s$^{-1}$ for Ross 128 and note that this vsin$i$ value is at the limit of what can be resolved from APOGEE spectra \citealt{Gilhool2018}. In all calculations we adopted a microturbulent velocity of $\xi$=1.00 km s$^{-1}$. 
The best fitted synthetic spectra were obtained by defining the pseudo-continuum of the observed spectra over each window analyzed, and the derived abundances are the value best-fitting the observed spectra from a $\chi$2 minimization method.

\begin{deluxetable}{lccccc}[h!]
\tabletypesize{\scriptsize}
\tablecaption{Atmospheric Parameters}
\tablewidth{0pt}
\tablehead{
\colhead{} &
\colhead{$T_{\rm eff}$} &
\colhead{log$g$} &
\colhead{$\xi$} &
\colhead{$[$Fe/H$]$} &
\colhead{$[$O/H$]$}
}
\startdata
MARCS 					& 3231$\pm$100	& 4.96$\pm$0.11	&	1.00	& +0.02	&	-0.03\\
PHOENIX					& 3223$\pm$100	& 4.89$\pm$0.11	&	1.00	& +0.03	&	-0.02\\
\cite{RojasAyala2012}	& 2986$\pm$50 		& ...				&	...		& +0.03	&	...	\\
\cite{Mann2015}			& 3192$\pm$64			& 5.08				&	...		& -0.02	&	...\\
\enddata
\tablewidth{0pt}	
\end{deluxetable}

\begin{figure*}
\figurenum{1}
\begin{center}
\includegraphics[angle=-90,width=0.95\linewidth,clip]{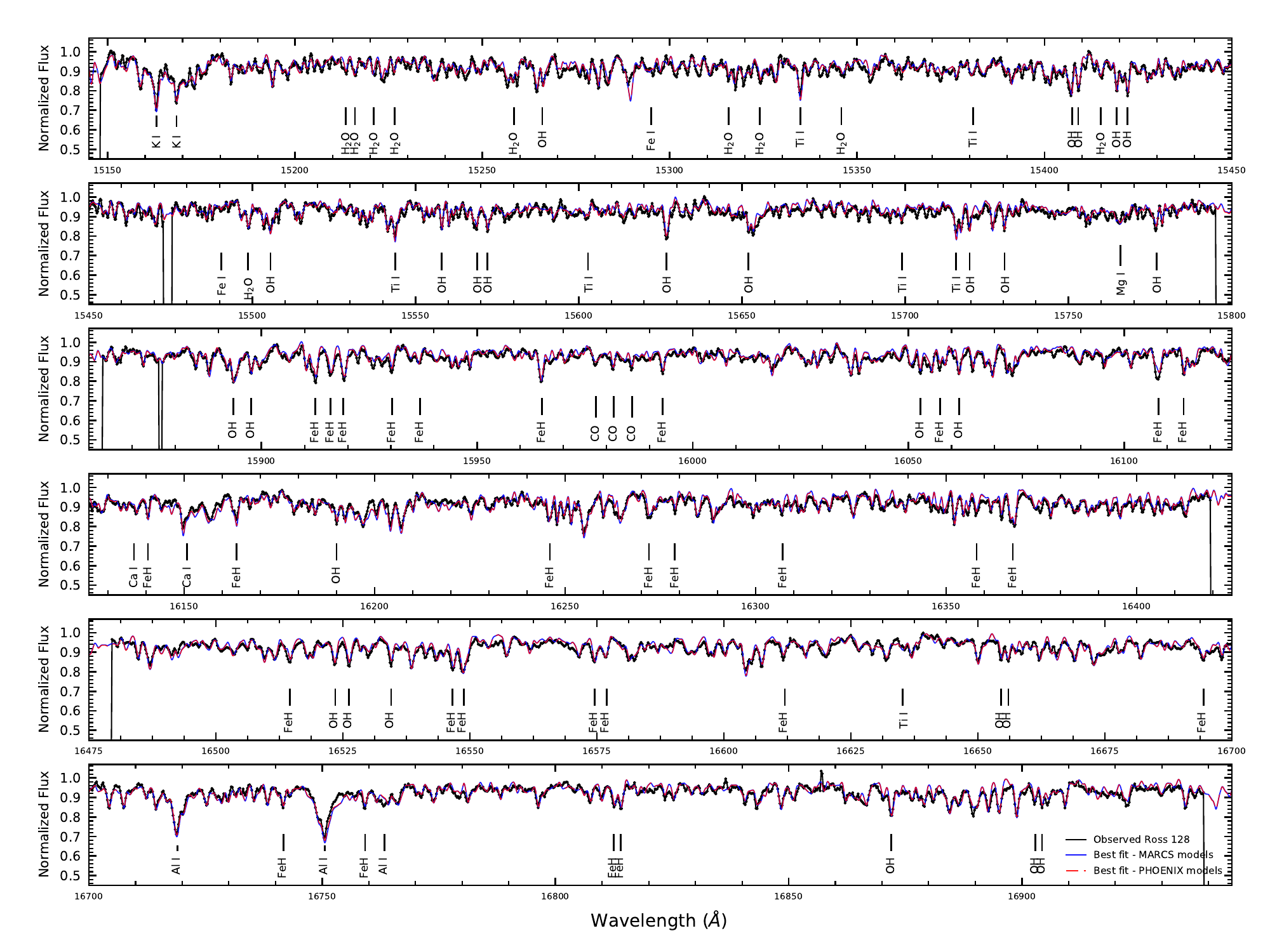}
\caption{The observed spectra (black) is shown superimposed over the best-fit MARCS (blue) and PHOENIX (red) synthetic spectra using the results from Tables 1 and 2. The lines or features used for the abundance analysis are labelled and indicated with black ticks.}
\end{center}
\label{fig1:fig1}
\end{figure*}

\subsection{Effective Temperature and Surface Gravity}

The effective temperature ($T_{\rm eff}$) and surface gravity (log$g$) values of Ross 128 were obtained using a similar methodology as discussed in \cite{Souto2017}. That previous study used the OH and H$_{2}$O lines as oxygen abundance indicators to define the effective temperature. Here we add Fe I and FeH as Fe abundance indicators and use the combination of oxygen and iron abundances from OH, H$_{2}$O, Fe I and FeH for deriving both the effective temperature and the surface gravity for Ross 128. 

Figure 2 shows the derived oxygen and iron abundances as functions of $T_{\rm eff}$ and log$g$ for both MARCS and PHOENIX model atmospheres. The effective temperature and log$g$ are defined by the agreement between the different abundance indicators (crossing points).
The derived effective temperature using the $T_{\rm eff}$--A(O) pair (top-left panel) is quite similar to the one from the $T_{\rm eff}$--A(Fe) pair (bottom-left panel), with a $T_{\rm eff}$ difference of only $\sim$5K; a similar effective temperature is derived using either family of model atmosphere. The same is true for log$g$, although with slightly larger differences ($\sim$0.02--0.08 dex). These results indicate that good consistency can be obtained between the different abundance indicators.

The stellar parameters and individual abundances derived from MARCS and PHOENIX models are very similar (Tables 1 and 2). We derive $T_{\rm eff}$= 3231/3223 $\pm$100K, log$g$= 4.96/4.89 $\pm$0.11 dex, and [Fe/H]= +0.03/+0.04 $\pm$0.09 (MARCS/PHOENIX).
The uncertainties in $T_{\rm eff}$ and log$g$ were estimated by allowing the Fe abundances from Fe I and FeH lines, and oxygen abundances from H$_{2}$O and OH lines to differ by 0.1 dex.  
The values from \cite{RojasAyala2012} displayed in Table 1 were obtained using the same techniques as \cite{Muirhead2014}. We adopt the $V$-$J$ and $r$-$J$ colors to determine the $T_{\rm eff}$ from \cite{Mann2015}.
The derived surface gravity (log$g$=4.96) agrees well with the log$g$ obtained from physical relation (log$g$=5.09) assuming $M_{\star}$=0.168$\pm$0.017$M_{\odot}$ from \cite{Mann2015}, $L$=0.00367$L_{\odot}$ from Gaia DR2 (\citealt{GaiaCollaborationDR2}), and the $T_{\rm eff}$ of this work using MARCS model, where $T_{\rm eff, \odot}$=5772K and log $g_{\odot}$=4.438 were adopted.
Table 2 presents the abundance results, the standard deviation of the mean abundances from the adopted lines (std), and the abundance uncertainties for each species ($\sigma$) to changes in the atmospheric parameters (computed following \citealt{Souto2017}).

\begin{deluxetable*}{lcccccccc}
\tabletypesize{\scriptsize}
\tablecaption{Individual Abundances}
\tablewidth{0pt}
\tablehead{
\colhead{} &
\colhead{} &
\colhead{MARCS} &
\colhead{} &
\colhead{} &
\colhead{PHOENIX} &
\colhead{} \\
\colhead{} &
\colhead{A(X)} &
\colhead{$[$X/H$]$} &
\colhead{std} &
\colhead{A(X)} &
\colhead{$[$X/H$]$} &
\colhead{std} &
\colhead{Number of Lines} &
\colhead{$\sigma$} 
}
\startdata
FeI 		& 7.48	& 0.03	& 0.02	& 7.49	& 0.04	& 0.02	&	2	&0.09\\
FeH 		& 7.45	& 0.00	& 0.07	& 7.46	& 0.01	& 0.08	&	30	&0.08\\
C 			& 8.41	& 0.02	& 0.01	& 8.41	& 0.02	& 0.01 	&	3	&0.02\\
OH 			& 8.62	& -0.04	& 0.02	& 8.63	& -0.03	& 0.02	&	29	&0.06\\
H$_{2}$O 	& 8.64	& -0.02	& 0.02	& 8.64	& -0.02	& 0.03	&	10	&0.08\\
Mg			& 7.43	& -0.10	& ...	& 7.48	& -0.05	& ...	&	1	&0.13\\
Al      	& 6.35	& -0.02	& ...	& 6.36	& -0.01	& ...	&	1	&0.10\\
K			& 5.03	& -0.05	& 0.03	& 5.06	& -0.02	& 0.03	&	2	&0.04\\
Ca			& 6.36	& -0.01	& 0.02	& 6.38	& 0.01	& 0.03	&	2	&0.03\\
Ti			& 4.73	& -0.17	& 0.21	& 4.77	& -0.13	& 0.20	&	6	&0.09\\
\enddata
\tablewidth{0pt}	
\end{deluxetable*}

\begin{figure*}
\figurenum{2}
\begin{center}
\includegraphics[width=0.99\linewidth]{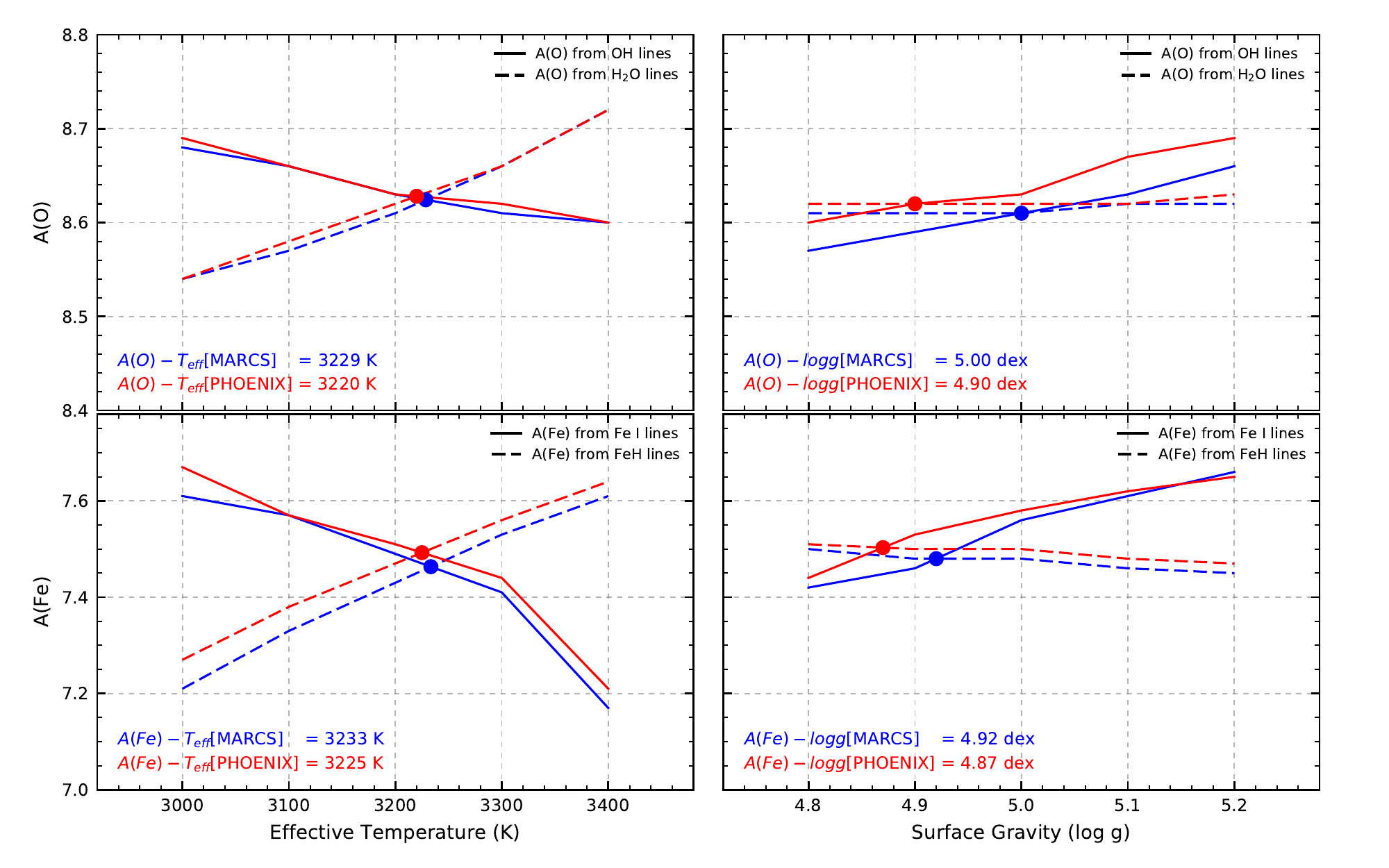}
\caption{Diagrams illustrating the derivation of atmospheric parameters.  The left panels display the $T_{\rm eff}$--A(O) and $T_{\rm eff}$--A(Fe) curves and the right panels the log$g$--A(O) and log$g$--A(Fe) pairs. 
In all panels, the derived abundances from Fe I and OH lines are shown as solid lines, while those from FeH and H$_{2}$O as dashed lines. The different colors (blue and red) indicate results derived from the adopted models, MARCS and PHOENIX respectively.}
\end{center}
\label{fig2:fig2}
\end{figure*}

\section{Discussion}

In the following discussion we adopt the abundance results derived with the MARCS model atmospheres, keeping in mind, however, that the results computed with PHOENIX models are very similar. 
The derived abundances of Ross 128 are close to solar for all elements, with the exception of Ti and to a lesser degree Mg (within the uncertainties). For Mg, we derive [Mg/H]=-0.10$\pm$0.13 dex from one neutral Mg I line ($\lambda$15765.842\AA{}), while for the other alpha-elements calcium and oxygen, we find just slightly subsolar abundances ([Ca/H]=-0.01$\pm$0.03 and [O/H]=-0.02$\pm$0.06). 
For Al we obtain [Al/H]=-0.02 from one Al I line at $\lambda$16763\AA{}; the other two Al I lines ($\lambda$16718 and 16750\AA{}) present in the APOGEE spectra of Ross 128 are too strong and were not used in our analysis (see also discussion in \citealt{Nordlander2017} for non-LTE effects in Al).
For Ti we obtain [Ti/H]=-0.17 from five Ti I lines with large scatter in the abundance results; the standard deviation of the mean is 0.21 dex indicating that the Ti abundance is uncertain to some degree. Non-LTE effects might be relevant for the Ti I H-band lines but these have not been investigated here nor included in our calculations. We also note that the results for Ti for the APOGEE DR14 red-giants present some issues as discussed in \cite{Souto2016}, and J\"onsson et al. (\textit{in preparation}).

\subsection{Star--Planet Connection}

One of the goals in studying individual abundances of exoplanet host stars is to try to infer the composition of the rocky exoplanets that orbit them. 
Certain abundance ratios (e.g. C/O and Mg/Si) play an important role in the chemistry of exoplanet formation in the disk and may also provide first-order information on the structure and mineralogy of the resulting rocky planets (e.g., \citealt{Bond2010}, \citealt{UnterbornPanero2017}). 
In this work we derive C/O=0.60$\pm$0.04 for Ross 128. While this is quite similar to solar (C/O=0.54, \citealt{Asplund2005}), it is below the value where refractory carbon is expected to condense and drastically affect the condensation temperatures of the refractory, rocky planet-building elements Mg, Si and Fe (C/O=0.8--1.0; \citealt{Lodders2003}). 
The stellar Mg/Si abundance ratio may affect the resulting silicate mineralogy of Ross 128b, although it is not possible to derive Si abundances in Ross 128 as the Si I lines become severely blended with H$_{2}$O or FeH. The Fe/Mg ratio, however, can affect the relative fraction of core-to-mantle and provide some broad constraints on the possible interior structure of Ross 128b.

While both mass and radius are not available for Ross 128b, we can estimate its radius given its observed minimum mass and assuming the stellar composition of the host star is a proxy for that of the planet. We calculate the range of radii possible for Ross 128b using the ExoPlex software package (\citealt{Unterborn2018}) for all masses above the minimum mass of Ross 128b (1.35M$_\oplus$; \citealt{Bonfils2017}). 
Models were run assuming a two-layer model with a liquid core and silicate mantle (no atmosphere). We increase the input mass until a likely radius of 1.5R$_\oplus$ was achieved, roughly the point where planets are not expected be gas-rich mini-Neptunes as opposed to rock and iron-dominated super-Earths (Figure 3, left). 
ExoPlex conserves the relative molar ratios for the dominant rocky planet-building elements assuming all Fe is present in the core with no light elements (e.g. Si, O, H) present in the core. Adopting our MARCS abundances we get the molar ratios Fe/Mg=1.12, Ca/Mg=0.09, and Al/Mg=0.08; where the Earth and solar values are 0.90/0.07/0.09 \citep{McDonough2003} and 0.83/0.06/0.08 \citep{Lodders2003}, respectively.

Changes in the relative planet's core size have a large effect on its bulk density (\citealt{Unterborn2016}) and we thus consider our Fe abundance uncertainty; 0.88$\leq$Fe/Mg$\leq$1.3.
We also varied Si/Mg between the extremes of 0.5 and 2 (Solar Si/Mg is 0.97). This range of Si/Mg represents an average difference in radius for a given input mass of roughly 0.05R$_\oplus$ (Figure 3, left panel). However, because the molar mass of MgO is 1/3 smaller than SiO$_2$ (our chosen mantle oxides), this range of Si/Mg represents core mass fractions between 0.44 and 0.27  (Earth is 0.33) for Si/Mg of 0.5 and 2, respectively. For this model, only if Ross 128b had a super-solar Si/Mg of 1.3 would it match the Earth in its core mass fraction, which equates to Ross 128 having a Si abundance of roughly solar (A(Si) = 7.54). 
That is to say while the molar ratio of Fe/Mg is conserved, the mass ratio of Fe/Mg in our model is not and depends on the chosen Mg/Si. This means that if Ross 128b's composition mimics that of its host star and a sub-Solar Si/Mg, it will have a larger relative core size than the Earth, despite having roughly solar iron abundance.
This is due to the relative ratio of Fe to Mg being greater than in the Sun and Earth, combined with the density of liquid Fe being greater than that of magnesium silicates (e.g. rock-dominated).
We note that the uncertainty in the derived radii is about the thickness of the red curve in Figure 3, left panel.

Our calculated Ross 128b radii -- under the assumption that the stellar chemical abundances are also those in the planet -- all lie below the 100\% rock composition curve of \cite{Zeng2016}, i.e., it contains a mixture of rock and Fe, with the relative amounts of each set by Fe/Mg.
Given our compositional constraints, these calculated radii represent the \textit{minimum} radius of Ross 128b in the absence of collisional stripping of silicate relative to iron (see also \citealt{Benz1988} and \citealt{Marcus2010}).
This means that the addition of light elements to the core or the presence of surface water or an atmosphere will only lower the planet density, causing a radius increase for the same mass. The Ross 128b relatively Fe-rich composition may affect its geodynamo (\citealt{Gaidos2010}, \citealt{Driscoll2011}), although the geodynamic and geochemical consequences of Fe-rich planets are not well explored.

\begin{figure*}
\figurenum{3}
\begin{center}
\includegraphics[width=0.39\linewidth]{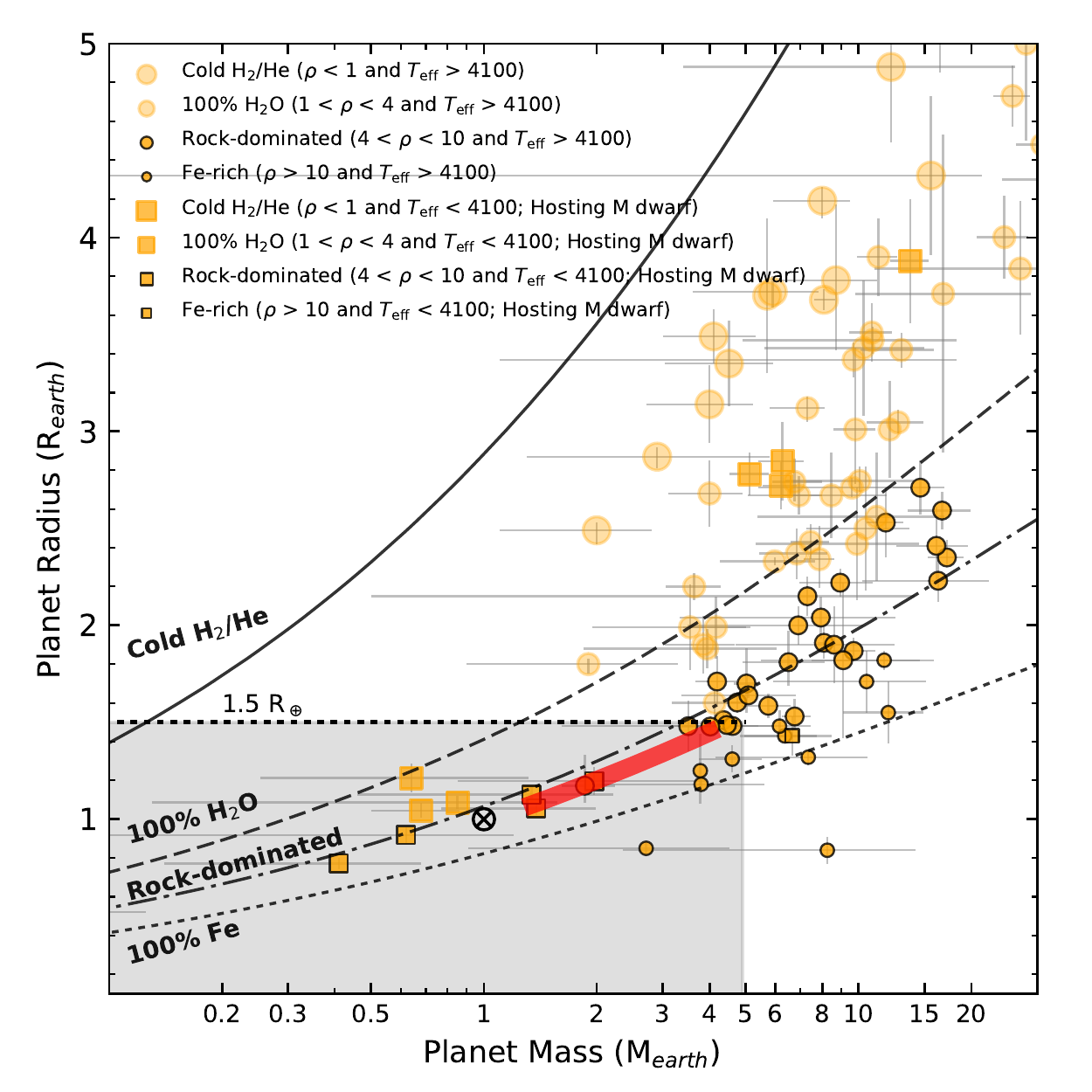}
\includegraphics[width=0.6\linewidth]{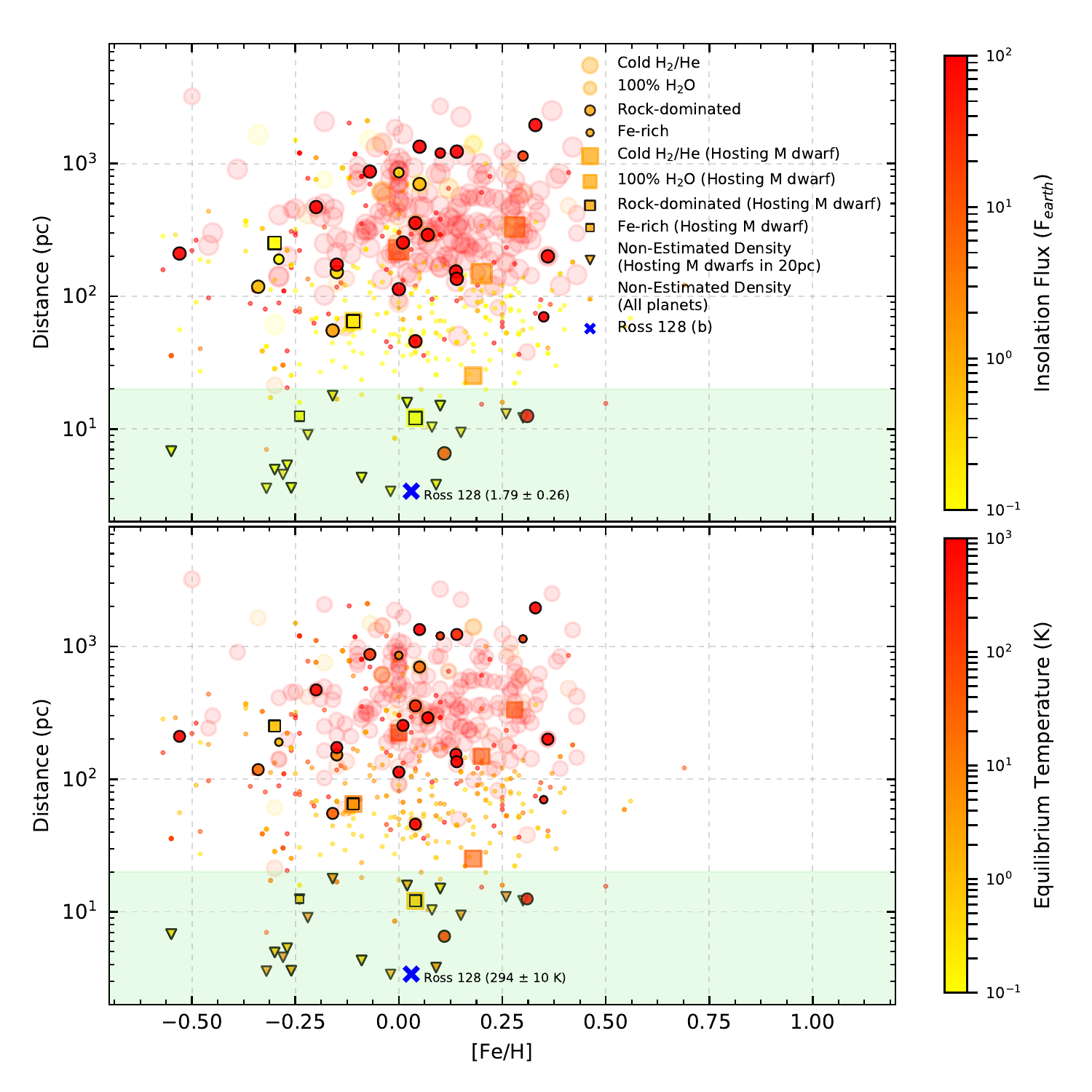}
\caption{Left panel: The exoplanet mass-radius diagram using the planetary composition based on \cite{Zeng2016} relations. Exoplanets around M dwarfs are shown with orange squares and the other stars hosting exoplanets are shown with orange circles. 
Right panel: The metallicity--distance diagram; in the top panel we plot the computed insolation flux at the planet's distance from the star and in the bottom panel we adopt the planet's equilibrium temperature in the color bar. 
The symbols follow the same notation from the left panel with the addition of upside down triangles for the M dwarfs that are closer than 20 pc from the Sun and host exoplanets without density measurements. We use dots to indicate the stars hosting exoplanets without density measured.}
\end{center}
\label{fig3:fig3}
\end{figure*}

With the results presented here, we can study some Ross 128b fundamental habitability parameters such as insolation flux (S$_{\rm Earth}$; flux of energy the exoplanet receives from its host star) and equilibrium temperature ($T_{\rm eq}$).
\cite{Bonfils2017} predict that Ross 128b should have an insolation flux (S$_{\rm Earth}$) of 1.38. 
Adopting our MARCS $T_{\rm eff}$, the stellar radius ($R_{\star}$=0.209$\pm$0.002 $R_{\odot}$; very similar to \citealt{Mann2015}) derived by us from the spectral energy distribution (\citealt{Stassun2017}) and adopting the \cite{Bonfils2017} semi-major axis (SMA=0.049 AU), we derive S$_{\rm Earth}$=1.79$\pm$0.26 for Ross 128b, which is consistent with \cite{Bonfils2017}. 
The $T_{\rm eq}$ of an exoplanet is a function of the stellar $T_{\rm eff}$, radius, the SMA of the planet and its albedo. 
We obtain for Ross 128b a $T_{\rm eq}$=294$\pm$10K for an Earth-like albedo (0.306). Adopting a Venus and Mars-like albedo (0.77 and 0.25), we derive a $T_{\rm eq}$ of 223$\pm$0.08 and 299$\pm$11K, respectively. 
Our results support the claim of \cite{Bonfils2017} that Ross 128b is a temperate exoplanet in the inner edge of the habitable zone. 
However, this is not to say that Ross 128b is a ``Exo-Earth.'' Geologic factors unexplored in \cite{Bonfils2017} such as the planet's likelihood to produce continental crust, the weathering rates of key nutrients into ocean basins or the presence of a long-term magnetic field could produce a planet decidedly not at all ``Earth-like'' or habitable due to differences in its composition and thermal history. 
Furthermore, other aspects of the M-dwarf's stellar activity and its effect on the retention of any atmosphere and potential habitability should be studied , although we find no evidence of activity in the Ross128 spectra.

In Figure 3 we use the NASA Exoplanet Archive to study the planetary density (left panel), the metallicity distribution of the host stars as well as the insolation flux and the planet equilibrium temperature as a function of the distance of the systems to the Sun (right panels).
We split the sample of known exoplanets into four groups based on the mass-radius-bulk composition curves from \cite{Zeng2016}. 
The planets with density ($\rho$)$<$1 were tagged as Cold H$_{2}$/He, with 1$<$$\rho$$<$4 have a predominant H$_{2}$O atmosphere, the rock-dominated exoplanet have 4$<$$\rho$$<$10, and those with $\rho$$>$10 are tagged as Fe-rich exoplanets;
we did not take density errors into account for these rough categories.
To give particular attention to the M dwarfs hosting exoplanets, we split the sample into stars with $T_{\rm eff}$$<$4100K (squared symbols) and $T_{\rm eff}$$>$4100K (circled symbols) using the same cuts in density.
In Figure 3 left panel, likely rocky-exoplanets are highlighted and the gray shadow region indicates the potential regime for Earth-like exoplanets (with radii from 0--1.5R$_{\earth}$ and masses from 0--5M$_{\earth}$).
The estimated radii for Ross 128b are displayed as a red solid line. The size of the red line represents the typical uncertainty in the estimated radii.

We show in the right panel of Figure 3 the metallicity ([Fe/H]) distance (pc) distribution of the host stars, with the exoplanet insolation flux (Figure 3 right top panel) and equilibrium temperature (Figure 3 right bottom panel) represented by color bars. 
We compute the exoplanet insolation flux and the equilibrium temperature using the $T_{\rm eff}$, R$_{\star}$, and SMA from the NASA Exoplanet Archive with an Earth-like albedo (30\%) for all exoplanets.
For the M dwarfs without $R_{\star}$ in the database, we adopt \cite{Mann2015} calibrations to determine the stellar radius.
With a green shadow, we highlight the region with $d$$\leq$20 pc, calling attention to the opportunity provided by the nearby M dwarf exoplanetary systems.
The symbols follow the same notation of Figure 3 left panel, with the addition of small dots systems where the exoplanet density is not measured.
The M dwarfs closer than 20 pc that host planets without measured densities are shown as upside down triangles. 
Ross 128b is also presented as blue crosses.
We leave the water and gaseous planets as background symbols.
From Figure 3, right top panel, the exoplanets with the lower degree of insolation flux orbit M dwarf stars (yellow squares and triangles). In contrast, the rocky exoplanets around solar-like stars tend to receive much more flux, generally 1000 times more than the Earth.\\

In summary, our precise spectroscopic atmospheric parameters and individual abundances have allowed us to use theoretical models to study the potential interior composition of Ross 128b. 
Assuming Ross 128b formed with the same composition as its host star we calculate its mineralogy, structure, and thus its mass. Our model assumes no atmosphere is present, however, because the addition of this layer (or the addition of light elements to the core) would decrease the density for a given radius, our calculated masses represent the \textit{maximum} mass of Ross 128b  in the absence of any mantle stripping due to large impacts. 
In this scenario, if Ross 128 is also depleted in Si relative to Solar we calculate Ross 128b would have a relatively larger core than the Earth, regardless of the mantle chemistry.
It is a likely scenario given the relative depletion of alpha-elements in Ross 128. 
The derived planetary parameters S$_{\rm Earth}$=1.79$\pm$0.26 and $T_{\rm eq}$=294$\pm$10K support the \cite{Bonfils2017} findings that Ross 128b is a temperate exoplanet in the inner edge of the habitable zone.

\acknowledgments

KC and VS acknowledge that their work here is supported, in part, by the National Aeronautics and Space Administration under Grant 16-XRP16\_2-0004, issued through the Astrophysics Division of the Science Mission Directorate. 
DAGH, OZ, and TM acknowledge support provided by the Spanish Ministry of Economy and Competitiveness (MINECO) under grant AYA-2017-88254-P. SRM acknowledges support from NSF grant AST-1616636. 
B.R-A acknowledges the support from CONICYT PAI/CONCURSO NACIONAL INSERCI\'ON EN LA ACADEMIA, CONVOCATORIA 2015 79150050.
H.J. acknowledges support from the Crafoord Foundation and Stiftelsen Olle Engkvist Byggm\"astare.
JKT acknowledge support for this work provided by NASA through Hubble Fellowship grant HST-HF2-51399.001 awarded by the Space Telescope Science Institute, which is operated by the Association of Universities for Research in Astronomy, Inc., for NASA, under contract NAS5-26555.

This research has made use of the NASA Exoplanet Archive, which is operated by the California Institute of Technology, under contract with the National Aeronautics and Space Administration under the Exoplanet Exploration Program.

Funding for the Sloan Digital Sky Survey IV has been provided by the Alfred P. Sloan Foundation, the U.S. Department of Energy Office of Science, and the Participating Institutions. SDSS-IV acknowledges
support and resources from the Center for High-Performance Computing at the University of Utah. The SDSS web site is www.sdss.org.

SDSS-IV is managed by the Astrophysical Research consortium for the 
Participating Institutions of the SDSS Collaboration including the 
Brazilian Participation Group, the Carnegie Institution for Science, 
Carnegie Mellon University, the Chilean Participation Group, the French Participation Group, Harvard-Smithsonian Center for Astrophysics, 
Instituto de Astrof\'isica de Canarias, The Johns Hopkins University, 
Kavli Institute for the Physics and Mathematics of the Universe (IPMU) / 
University of Tokyo, Lawrence Berkeley National Laboratory, 
Leibniz Institut f\"ur Astrophysik Potsdam (AIP),  
Max-Planck-Institut f\"ur Astronomie (MPIA Heidelberg), 
Max-Planck-Institut f\"ur Astrophysik (MPA Garching), 
Max-Planck-Institut f\"ur Extraterrestrische Physik (MPE), 
National Astronomical Observatory of China, New Mexico State University, 
New York University, University of Notre Dame, 
Observat\'orio Nacional / MCTI, The Ohio State University, 
Pennsylvania State University, Shanghai Astronomical Observatory, 
United Kingdom Participation Group,
Universidad Nacional Aut\'onoma de M\'exico, University of Arizona, 
University of Colorado Boulder, University of Oxford, University of Portsmouth, 
University of Utah, University of Virginia, University of Washington, University of Wisconsin, 
Vanderbilt University, and Yale University.

\facility {Sloan, NASA Exoplanet Archive}

\software{ExoPlex (\citealt{Unterborn2018}), Turbospectrum (\citealt{AlvarezPLez1998}, \citealt{Plez2012}), MARCS (\citealt{Gustafsson2008}), PHOENIX BT-Settl (\citealt{Allard2013})}

\end{document}